# Tone Biased MMR Text Summarization

[1]Mayank Chaudhari, [2]Aakash Nelson Mattukoyya

[1,2]Post Graduate Student
[1,2]Department of Computer Science & Information Systems
[1,2]BITS-Pilani, K. K. Birla Goa Campus, Goa, India

___

*Abstract:* Text summarization is an interesting area for researchers to develop new techniques to provide human like summaries for vast amounts of information. Summarization techniques tend to focus on providing accurate representation of content; and often the tone of the content is ignored. Tone of the content sets a baseline for how a reader perceives the content. As such being able to generate summary with tone that is appropriate for the reader is important.

In our work we implement Maximal Marginal Relevance [MMR] based multi-document text summarization and propose a naïve model to change tone of the summarization by setting a bias to specific set of words and restricting other words in the summarization output. This bias towards a specified set of words produces a summary whose tone is same as tone of specified words.

___

## I. INTRODUCTION

With vast amounts of information being generated every day, automated text summarization is used to represent such vast information in compact form. There are many techniques that have been developed over recent times to improve the accuracy of summary and provide summaries that are 'human like'. Various features of sentences are used to rank the sentences which should be included in the summary. The top ranked sentences are refined and reordered to form a coherent summary. Feature based ranking means that we try improving the query relevance of the sentences selected for summarization but this in turn might increase redundancy in summary as many query relevant sentences may have similar content. In our work we implemented Maximal Marginal Relevance [MMR] based text multi-document summarization which along with sentence ranking considers the novelty of the sentence there by reducing the redundancy in final summarization.

Current summarization techniques try to make sure that all information in content is accurately represented in the summary. Such summarization gives us accurate representation of data, however, the summarization isn't 'human like'. Daily life human summaries usually tend to be customized for reader by using bias in the tone of summarization. We propose a naïve model for biasing the tone of summary by using a set of words which have defined polarity tag to decide whether to include or discard the sentence in the summary.

The rest of the paper is organized as follows – Section II gives an overview of current work done in the proposed area. Section III discusses our approach and implementation of MMR multi-document text summarization and Naïve tone biasing. Section IV briefly discusses the applications of tone biasing. Section V we discuss the results and observations obtained by implementing our approach. Section VI describes future scope of our work and conclusions.

## II. LITERATURE SURVEY

The ability to get deeper insights without having to manually read through huge amount of data has fueled research in field of text summarization. Carbonell and Goldstein [1] proposed maximal marginal relevance and discussed the MMR based text summarization in detail. Long et al., [2] discuss ways to apply learning models for optimizing diversity evaluation measure in training and proposes a novel modelling approach Perceptron. Xie and Liu [3] compare the various knowledge-based similarity measures that can be used with MMR in summarization of large corpus of meeting recording according to their experimental rogue scores.

Radev et al., [4] present a multi-document summarizer MEAD which uses cluster centroids produced by topic detection and tracking to generate summaries. Goldstein et al., [5] propose a new approach based on domain independent techniques for multi-document text summarization which has few operations based on single document text summarization. Yulita and Pribadi [6] implemented [1] with simple modifications such as using TF-IDF-DF for ranking sentences. Yadav and Chatterjee [7] discuss the application of sentiment analysis for text summarization using various summary techniques and compared them. Gupta et al., [8] surveyed the existing text summarization methods which integrate well with ML and AI techniques for sentiment analysis for online product reviews.

We see that most works in MMR text summarization work on augmenting MMR with other algorithms to improve accuracy.

## III. APPROACH

Our approach starts by data preprocessing then using MMR to fetch relevant sentences and remove redundancy to improve novelty. Now, we have a content accurate novel text summary of the document, but the tone of the summary just reflects the tone of the content. We bias the tone of the generated summary using proposed naïve approach.

Following subsections describe the above-mentioned approach step by step –

### A. Preprocessing

Firstly, we must clean our data set to be able to run our algorithms efficiently and remove any unwanted/unnecessary data. In this step we scan through all the documents to remove stop words and XML tags which are unnecessary while processing. The sentences are classified into separate entities as our next steps will process text as sentences. Lastly, we reduce the words of each sentence into its stem word using the Porter Stemming algorithm.

### B. Sentence ranking using TF-IDF values

TF-IDF values give us the relevance of a sentence with respect to the query vector so that we can pick the top-k relevant sentences. The query vector is generated by finding most frequent words in the document that reflect the subject of document. Since our query vector is based on frequent words, the TF-IDF sentence ranking returns the sentences that are most close to the general summary of the document by using cosine similarity.

Equation for calculation of TF-IDF is given below –

$$w_{i,j} = tf_{i,j} \, x \log\left(\frac{N}{df_i}\right)$$

$w_{i,j}$ is the weight of the word 'i' in the sentence 'j'. $tf_{i,j}$ (term frequency) is the term frequency of word 'i' in sentence 'j'. $\log(N/df_i)$ is the equation of Inverse Document Frequency (IDF), N is the number of the total sentences. $df_i$ (document frequency) is the number of sentences which contain the word 'i'.

We can also vary the TF-IDF method to use TF-IDF-DF as used in [6], or use other similarity measures like Pearson's coefficient, but the main objective of our work is to explore tone biasing and evaluate a naïve approach.

### C. Maximal Marginal Relevance

The Maximal Marginal Relevance [MMR] technique tries to reduce the redundancy while maintaining relevance to the query when reordering sentences. We calculate the relevance of the query and sentences using cosine similarity then we calculate the similarity of sentences among themselves and remove the similar sentences to remove redundancy. The below equation shows the MMR

$$MMR(S_i) = \lambda \cdot Sim_1(S_i, Q) - (1 - \lambda) \cdot \max Sim_2(S_i, S')$$

Sim1 as explained in last subsection is rank of sentence in terms of best word query. Sim2 is the cosine similarity of current sentence among the list of top-n sentences S` that we get from Sim1. '$\lambda$' is the tunable parameter which allows the user to tune the MMR equation. $\lambda$ value ranges between 0 to 1, where 0 indicated maximum similarity and 1 indicates maximal diversity. MMR works iteratively to get the best possible non- redundant summary. MMR process stops when MMR(Si) becomes less than zero.

We start by varying $\lambda$ from 0.3 to 0.9 and observe that we get best accuracy when $\lambda$ is between 0.5 and 0.7 for given dataset. We used DUC2001 dataset for performing the text summarization.

### D. Tone Biasing

As discussed in the introduction we propose a naïve approach for biasing the tone of summary. We use TextBlob library in python which provides functions to compute the polarity of sentences and tag them as positive, negative and neutral. Polarity of the sentence is calculated by summing the polarity of words in the sentence. TextBlob internally does stemming and lemmatization to provide accurate polarity information.

We use this method to analyze the polarity of the sentence retrieved after TF-IDF sentence ranking, then if the sentence has negative polarity we discard the sentence. Finally, only the sentences with either neutral or positive polarity are populated in the top-n list passed to MMR $Sim_2$. This results in MMR producing positive summaries due to the positive bias. We can also flip the tone by simply discarding positive polarity sentences instead of negative polarity sentences.

Polarity is context sensitive and is referred abstractly in the paper; it needs to be explicitly defined by the users according to their use case. Users can learn the polarity of the words and define the tone according to the context by tweaking their classifiers used to assign polarity tags in TextBlob.

We've developed this naïve approach as we are exploring the tone biasing approach. We can make this approach more sophisticated and robust by augmenting other text summarization techniques and polarity tagging techniques, we discuss a few such approaches in Future work section.

## IV. APPLICATIONS

System generated summaries are generally consistent and content accurate. While these are desirable properties, in the quest to make summaries as human-like as possible we need bias the summary according to our audience. Tone biasing has lot of applications where text summaries should be modified to convey the same message in different flavor. We could think of many applications of tone biasing, here we present two such examples.

### (i) Censoring content in a graceful manner

If a similar content should be displayed to different age groups, it might be beneficial to change the tone of summary accordingly. Apart from just adding removing sentences based on polarity we can work on an Natural Language Processing approach to use similar words with less negativity as replacement for existing words to improve the polarity of sentence towards required direction.

*(ii) Summarizing product/service reviews*

We can summarize reviews positively or critically for different audiences. Positive summary could be provided to prospective buyers, while critical summary could be provided to backend teams as feedback to improve service.

We can also extend this approach by having a multiclass summary instead of just negative or positive summary. The multiple classes would depend on the context and subject of the documents we're trying to summarize.

## V. RESULT

We implemented the MMR approach described by Carbonell and Goldstein [1], and experimental results were observed to be lower than the original paper as the original paper uses normalized recall and f-score. We evaluated our implementation using the DUC2001 dataset which contains news articles on various topics. DUC2001 dataset also provides 100 words, 200 words and 400 words summary of topics prepared by humans as a benchmark to evaluate the accuracy of our approaches. We use rogue score which is the combination of {Recall, Precision and F-Score} to measure the accuracy of our approach. Also, we've implemented our naïve tone biasing approach for biasing the text and compared it with polarity of non-biased summary.

We present our observations below –

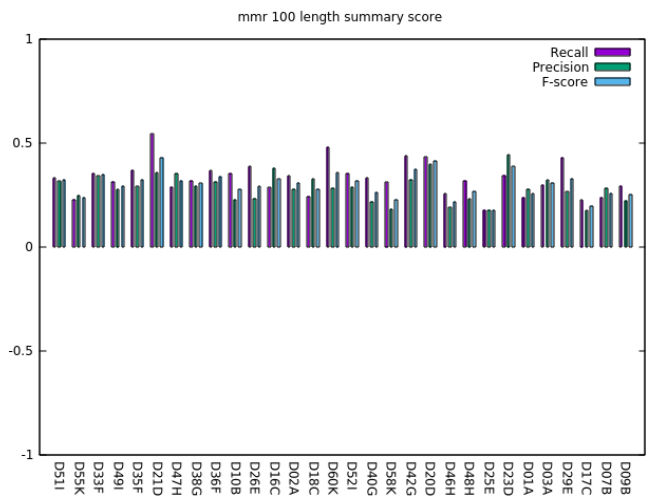

*Figure 1 Rouge score of MMR summary of length 100 words*

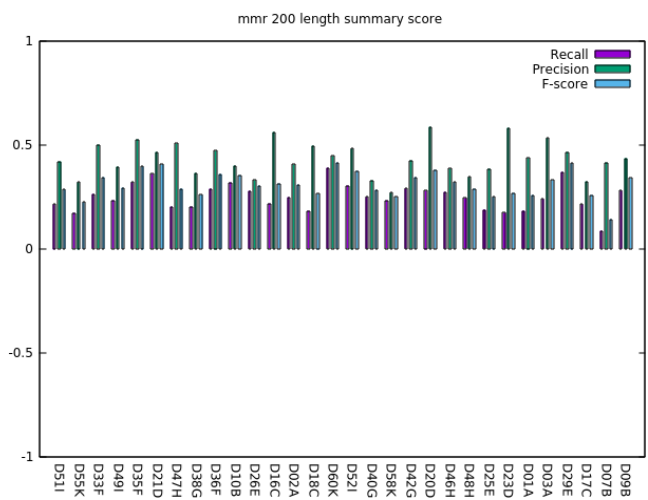

*Figure 2 Rouge score of MMR summary of length 200 words*

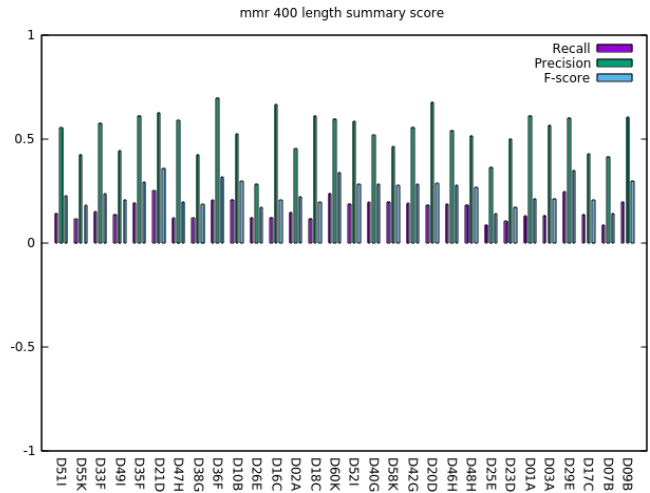

*Figure 3 Rouge score of MMR summary of length 400 words*

Figures [1][2][3] show plot of the rouge scores for document cluster of 30 documents of DUC2001 dataset. We see that MMR multi document text summarization technique provides an average recall of 34.8%, average precision of 45.2% and average F-score of 41.9%. Average rouge score for our implementation of MMR summarization would be {34.8, 45.2, 41.9}. This is lower when compared to the average rouge score of original paper which was {44.8, 45.9, 45.83}, this is as the original paper uses normalized recall and F-score to provide better accuracy and use TF-IDF-ID instead of TF-IDF for sentence ranking. In best case scenario our approach has the rogue score of {60.2, 71.9, 47.8} this is similar to the original paper's rogue score.

We also observe that as the length of summary increases from 100 words to 400 words the recall value and f-score decrease while precision increases. Recall value is the total number of correct words that are present in summary versus total correct words, as the length of summary increases the probability of fetching incorrect words increases and as the number of sentences we pick to calculate MMR remains same the recall value is impacted. The same goes for f-score. However, precision is the total number of correct words in the summary versus total words in the summary, this means that when summary length increases most correct words could be included in summary, this leads to increase in precision value.

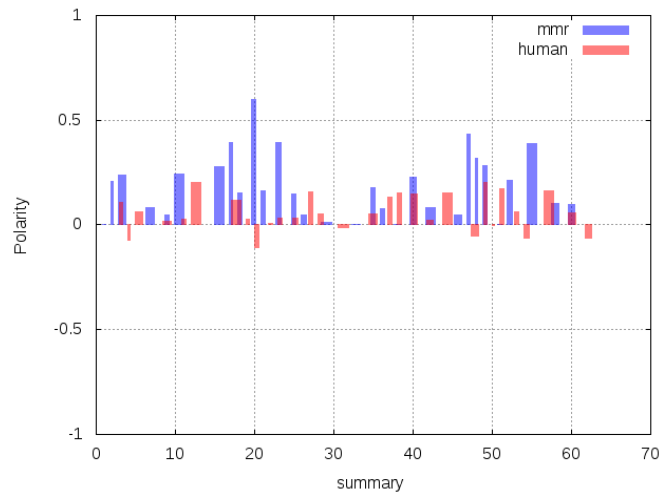

*Figure 4 Polarity of document clusters with and without tone biasing*

Figure [4] is the plot comparing the polarity of sentences in MMR with Naïve tone biasing approach indicated in blue color and human summary without any tone bias dataset denoted in red color. MMR was implemented using *λ*=0.7 and summary chosen for evaluation was 400-word length summary provided in DUC2001 dataset. We see that few documents have negative polarity in human summary and MMR with naïve tone biasing successfully changes the polarity of all documents to positive polarity. We can observe that few document clusters find have negative polarity higher than positive polarity, this indicates that the document has high negative polarity and converting such document to positive polarity meant dropping lot of negative sentences which results in loss of information.

## VI. CONCLUSION AND FUTURE WORK

We observed that MMR based multi-document text summarization provides good accuracy and is superior to other approaches removing redundant information in summary. Naïve tone biasing approach works well but might result in information loss when the tones we don't need are too high in the document. Such loss of information can be mitigated by rephrasing sentences with required tone instead of discarding them, this is part of a larger NLP problem and there is future scope for us to work this area.

Naïve tone biasing approach does binary biasing, i.e. either positive bias or negative bias and meaning of positive and negative might vary with the subject. As such instead of having a static lexicon of polarity scores of words we need develop a model to dynamically generate a lexicon with polarities according the subject. Such a model would give us better accuracy as the lexicon is customized to the subject/topic in point. Also, in some subjects/topics it might be better to have multiclass biasing instead of having only two classes. We need to develop a model to identify different classes of information and construct a polarity lexicon.